\documentclass[12pt]{article}
\usepackage{latexsym}
\usepackage{epsfig,amssymb,euscript}
\usepackage{amsmath}
\usepackage{array,calc,epsfig}
\usepackage{graphicx}
\usepackage{dcolumn}
\usepackage{bm}

\def\RR{{\rm I\kern-1.6pt {\rm R}}}

\newcommand{\rc}{\nonumber\\}
\newcommand{\bear}{\begin{eqnarray}}
\newcommand{\eear}{\end{eqnarray}}

\newcommand{\be}{\begin{equation}}
\newcommand{\ee}{\end{equation}}
\newcommand{\beq}{\begin{equation}}
\newcommand{\eeq}{\end{equation}}
\newcommand{\bea}{\begin{eqnarray}}
\newcommand{\eea}{\end{eqnarray}}

\newcommand{\ba}{\begin{eqnarray}}
\newcommand{\ea}{\end{eqnarray}}

\def\lbldef#1#2{\expandafter\gdef\csname #1\endcsname {#2}}

\def\href#1#2{#2}

\newcommand{\ber}{\begin{eqnarray}}
\newcommand{\eer}{\end{eqnarray}}
\newcommand{\beqar}{\begin{eqnarray}}

\newcommand{\eeqar}{\end{eqnarray}}

\newcommand{\dsl}
  {\kern.06em\hbox{\raise.15ex\hbox{$/$}\kern-.56em\hbox{$\partial$}}}

\newcommand{\eeqarr}{\end{eqnarray}}
\newcommand{\ZZ}{{\rm \kern 0.275em Z \kern -0.92em Z}\;}

 \def\vol{{\hbox{\rm Vol}}}

\def\CC{{\mathchoice
{\rm C\mkern-8mu\vrule height1.45ex depth-.05ex
width.05em\mkern9mu\kern-.05em}
{\rm C\mkern-8mu\vrule height1.45ex depth-.05ex
width.05em\mkern9mu\kern-.05em}
{\rm C\mkern-8mu\vrule height1ex depth-.07ex
width.035em\mkern9mu\kern-.035em}
{\rm C\mkern-8mu\vrule height.65ex depth-.1ex
width.025em\mkern8mu\kern-.025em}}}
                                                                                                    
\def\RR{{\rm I\kern-1.6pt {\rm R}}}

\def\ZZ{{\rm Z}\kern-3.8pt {\rm Z} \kern2pt}
\def\IB{\relax{\rm I\kern-.18em B}}
\def\ID{\relax{\rm I\kern-.18em D}}
\def\II{\relax{\rm I\kern-.18em I}}
\def\IP{\relax{\rm I\kern-.18em P}}

\def\to{\rightarrow}

\def\to{\rightarrow}

  \def\th{\theta}                  

\def\6{\partial}

\newfont{\namefont}{cmr10}
\newfont{\addfont}{cmti7 scaled 1440}
\newfont{\boldmathfont}{cmbx10}
\newfont{\headfontb}{cmbx10 scaled 1728}
\renewcommand{\theequation}{{\rm\thesection.\arabic{equation}}}
\begin{document}
\baselineskip=15.5pt
\pagestyle{plain}
\setcounter{page}{1}

\begin{titlepage}

\vskip -.5cm
\leftline{\tt hep-th/0702225}
\vskip -.8cm

\rightline{\small{\tt UB-ECM-PF 07/03}}

\rightline{\small{\tt CECS-PHY-07/04}}
\vskip 1.3 cm


\begin{center}
{\Large{\bf Holography and Unquenched Quark-Gluon Plasmas}}
\vskip 1.2cm
\vspace{20pt}

{\large G. Bertoldi $^{a}$, F. Bigazzi $^{b}$, A. L. Cotrone $^{c}$, J. D.
Edelstein $^{d,e}$}

\vspace{20pt}

\textit{$(a)$  Department of Physics, Swansea University, 
Swansea, SA28PP, UK.}\\
\textit{(b)  Physique Th\'eorique et Math\'ematique and International Solvay
Institutes, \\Universit\'e Libre de Bruxelles, C.P. 231, B-1050
Bruxelles, Belgium.}\\
\textit{(c) Departament ECM, Facultat de F{\'\i}sica, Universitat de
Barcelona and\\ Institut de F{\'\i}sica d'Altes Energies,
Diagonal 647, E-08028 Barcelona, Spain.}\\
\textit{$(d)$ Departamento de F{\'\i}sica de Part{\'\i}culas and IGFAE, \\
Universidade de Santiago de Compostela, E-15782 Santiago de
Compostela, Spain.}\\ 
\textit{$(e)$  Centro de Estudios Cient{\'\i}ficos (CECS), Casilla 1469,
  Valdivia, Chile.}\\

\end{center}

\vspace{12pt}

\begin{center}
\textbf{Abstract}
\end{center}

\vspace{4pt} {\small \noindent
We employ the string/gauge theory correspondence to study properties
of strongly coupled quark-gluon plasmas in thermal gauge theories with a
large number of colors and flavors. In particular, we analyze non-critical
string duals of conformal (S)QCD, as well as ten dimensional wrapped
fivebrane duals of SQCD--like theories. We study general properties of
the dual plasmas, including the drag force exerted on a probe quark and
the jet quenching parameter. We find that these plasma observables depend
on the number of colors and flavors in the ``QCD dual''; in
particular, we find that the jet quenching parameter increases
linearly with $N_{f}/N_{c}$ at leading order in the probe limit. In the ten
dimensional case we find a non trivial drag coefficient but a vanishing
jet quenching parameter. We comment on the relation of this result with
total screening and argue that the same features are shared by all known
plasmas dual to fivebranes in ten dimensions. We also construct new D5
black hole solutions with spherical horizon and show that they exhibit
the same features.}

\vfill
\vskip 5.mm
\hrule width 5.cm
\vskip 2.mm
{\small
\noindent g.bertoldi@swansea.ac.uk, fbigazzi@ulb.ac.be, cotrone@ecm.ub.es,
jedels@usc.es}
\noindent
\end{titlepage}
\newpage

\section{Introduction}

The experimental data collected at RHIC \cite{RHIC}, where heavy ions
collide at very high energies, indicate that a new state of matter is
formed during the collisions: the QCD quark gluon plasma (QGP). It partially
behaves like a fluid with very low shear viscosity and, contrary to the
common theoretical expectations, present evidence supports the picture of
a strongly interacting medium \cite{shur}. Some of its observed features
can be fitted with predictions from lattice gauge theory, but this cannot
be used to describe its real time dynamical properties.

A powerful tool to study strongly interacting gauge theories, both at 
finite and at zero temperature, is the string/gauge theory correspondence
\cite{malda}. At present it has been used to study the properties of
plasmas where gluons and adjoint matter fields (but no quarks) interact.
The high temperature gauge theory at strong 't Hooft coupling $g^2_{YM}
N_c \gg 1$, and in the limit of large number of colors $N_c \gg 1$, can
be described by a dual classical supergravity model with background metric
containing a black hole. The features of the metric depend on those of the
dual field theory. The master example is provided by the thermal version
of ${\cal N}=4$ Supersymmetric Yang-Mills (SYM) theory. The dual gravity
background is $AdS_5\times S^5$ with a large black hole in the center
of the space. The above thermal field theory differs from thermal QCD
in many aspects, one of the most evident being its lack of dynamical
quarks. Nevertheless, the ${\cal N}=4$ SYM plasma is non supersymmetric
and non confining as the QCD one, and it is tempting to compare the
$AdS/CFT$ predictions on ${\cal N}=4$ SYM and the extrapolated data
from QCD in order to understand how macroscopic observables are sensitive
to the microscopic details of these two theories. 

The string/gauge correspondence has been employed to evaluate the shear
viscosity of the ${\cal N}=4$ plasma \cite{pss} some years ago. This was
a major success since the resulting value seems to obey a universal law
and, furthermore, it is well compatible with the values for the QGP found
at RHIC. This heavily supports the use of $AdS/CFT$ to describe this physics.
More recently, a lot of explicit calculations of real time dynamical
properties, such as the drag force experienced by a quark moving through
the plasma \cite{herzog1,gubser,casal}, the jet quenching parameter measuring
the average energy loss by the parton per mean free path \cite{liu}, the
screening length  of a moving color charge inside the plasma
\cite{liu2,chergu},
and the rate of photoproduction \cite{photo} were given. Generalizations
of these results in different directions were pursued, respectively, in
\cite{dragforce}, \cite{jetquenching}, \cite{screeninglength} and
\cite{photoproduction}.
See, also, \cite{herzog2,hep-ph/0608274,hep-ph/0609254,antonyan,liu3}.

Despite some remarkable and unexpected numerical similarities between
observables evaluated in the extrapolation to $N_c=3$ of large $N_c$,
${\cal N}=4$ thermal SYM theory and the same observables deduced for
the QCD quark--gluon plasma from the RHIC data, there are several
reasons to be cautious. Of primary importance, it is mandatory to put
the string/gauge theory correspondence at work for field theories 
closer and closer to real world QCD. In this sense, an unavoidable
step to be able to properly talk about a QGP consists in dealing
with gauge theories coupled to $N_f$ flavors of dynamical quarks.
The first order effects in $N_f/N_c$ of the inclusion of flavors in
the thermal string/gauge theory correspondence have been recently 
studied in \cite{babi,myers,mateosf,Myers}. In this paper, we aim at
considering the full, unquenched influence of dynamical quarks on
the plasmas, by analyzing known supergravity solutions encoding
these effects.

In Section \ref{noncrit}, we shall study plasmas having a non-critical
string theory dual description. The relevant gravity backgrounds we
consider are  $AdS_{5}$ and $AdS_{5}\times S^{1}$ black hole solutions
found in \cite{cps}. They are the non-extremal deformations of the zero
temperature solutions found, respectively, in \cite{paris} and \cite{km}.
These are conjectured to be dual to conformal QCD and ${\cal N}=1$ SQCD
in the Seiberg conformal window, in the so-called Veneziano limit
\cite{veneziano}, where $N_f, N_c \rightarrow \infty$ but $\rho \equiv
N_f/N_c$ is kept finite. The gravity solutions are strongly curved and
$\alpha'$ corrections are expected to be relevant. Nevertheless, it is
worth exploring their properties, as they give a hint on the qualitative
behavior of plasma observables in relation with $N_f$. The holographic
expressions for plasma properties like the screening length, the drag
force coefficient, the jet quenching parameter and the diffusion
coefficient for heavy quarks, can be immediately taken from the $AdS$
critical cases, as they are not sensitive to the transverse directions.
Interestingly enough, when expressed in terms of gauge theory quantities,
these observables turn out to be independent on the number of colors and
flavors in the ``SQCD'' model. This is not so for the ``QCD'' model,
where, apart for the screening length which remains insensitive to
the degrees of freedom of the plasma, there is an explicit dependence
on $N_c, N_f$. In particular, the drag force coefficient and the jet
quenching parameter, which are the quantities on which we will mainly
focus, are monotonic growing functions of $\rho$.
This could be connected to a similar behavior of the entropy
density. This qualitative behavior would reinstate the discussion about
the use of the jet quenching parameter as a direct measure of the
temperature of the plasma, since it would explicitly depend on the
number of effective degrees of freedom. Finally, we make the (trivial)
observation that the shear viscosity to entropy density ratio is still
$1/4\pi$ in these backgrounds, in the gravity approximation.

In Section \ref{fivebranes}, we analyze a ten dimensional one-parameter
family of black hole solutions found in \cite{cnp} and conjectured to
be dual to the high temperature regime of an ${\cal N}=1$ SQCD-like
theory with $N_f = 2 N_c$ flavors and adjoint matter. The backgrounds
describe the low energy dynamics of a stack of a large number of both
wrapped ``color'' D5--branes and smeared ``flavor'' branes in the
Veneziano limit. We study again the energy loss of a probe moving
into the dual plasmas by evaluating the drag force and the jet quenching
parameter. Moreover, we study the screening properties of the plasmas.
We find that the drag force has analogous functional form as the one
evaluated for the ${\cal N} = 4$ SYM plasma, that the jet quenching
parameter vanishes and that probe quarks and anti-quarks are always
screened. These results seem related to the critical thermodynamical
properties of the model, in particular the fact that the black hole
temperature, and so the temperature of the gauge theory, coincides with
the Hagedorn temperature of Little String Theory (LST). We argue that
the same properties are common to all known models, with or without
flavors (as the thermal Maldacena--N\'u\~nez solution
\cite{mn,Buchel:2001qi,gtv}), having an UV LST completion. In
particular, as shown explicitly in appendix \ref{gtv} for the
backgrounds found in \cite{gtv}, the LST plasmas are well expected
to totally screen quark-antiquark interactions. The screening property
of these plasmas should be responsible for the vanishing of the jet
quenching parameter.

We also present in appendix \ref{spherical} a new family of black hole
solutions which arise as compactifications on $S^3$ of the $N_f = 2 N_c$
configuration of \cite{cnp}. Despite the inclusion of a new scale in the
theory (the compactification radius), that naively might have been
expected to modify the temperature away from the Hagedorn point, their
holographic duals have the same thermodynamic problems and screening
properties of the other LST plasmas.

We conclude in Section \ref{conclusions} with a brief discussion. 

\section{QGP and non-critical holography}
\label{noncrit}

Non-critical string duals of 4d gauge theories with large $N_c, N_f$
both at zero and at high temperature have appeared in the literature in
the last few years. In this section we investigate properties of the
quark--gluon plasmas in these contexts by using the string/gauge theory
correspondence, though the gravity solutions at our disposal are generically
strongly curved and $\alpha'$ corrections are not subleading. Our optimistic
prejudice, driven by the unexpected success of bottom--up approaches such
as the so-called ``$AdS$/QCD'' duality \cite{katz,pomarol}, is that the
$\alpha'$ corrections may be not very large, and that the non-critical
solutions should provide at least qualitative information on the dual
field theories.

The explicit backgrounds we will consider are an $AdS_{5}$ black hole
solution which was proposed as the non-critical string dual of the thermal
version of conformal QCD and an $AdS_{5}\times S^{1}$ black hole solution
conjectured to be dual to thermal ${\cal N}=1$ SQCD in the Seiberg conformal
window \cite{cps}. The zero temperature models were constructed in
\cite{paris} and \cite{km} respectively. In both models the color
degrees of freedom are introduced via $N_{c}$ D3--brane sources and the
backreaction of $N_{f}$ ``flavor'' branes on the background is taken
into account. The flavor branes are spacetime filling brane-antibrane
pairs, \footnote{More precisely, in the six-dimensional model a
brane-antibrane pair is just a single D5--brane doubly wrapped on the
cigar \cite{prezas}.} so to reproduce the classical $U(N_f) \times
U(N_f)$ flavor symmetry expected in the gauge duals with massless
fundamental matter (see \cite{ckp} for a very interesting analysis of
this kind of setting).

In order to study the dynamics of hard probes in the quark--gluon plasmas
of these theories, we will make the assumption that the mass of the probes
is related to the radial distance of a flavor brane from the center of the
space, as it happens in the critical $AdS_5 \times S^5$ example, at least
in some effective way. \footnote{Even in the model of \cite{ckp}, where also
the branes supporting massive quarks are space-time filling, one expects a
probe quark to be described by a string attached to the brane at finite
distance, proportional to its mass, from the bottom of the space.} In this
case the general results on the drag force and jet quenching parameter
obtained for $AdS_{5}$ black holes in the critical case,
extend in a straightforward way to the non-critical setup. Indeed, the
general form of the background string frame metric in the non-critical
cases under study is
\be
ds^{2}= \left( \frac{u}{R} \right)^{2} \left[ \left(1 - \frac{u^4_H}{u^4}
\right) dt^{2} + dx_{i}dx_{i} \right] + (R u)^{2} \frac{du^{2}}{u^{4} -
u_{H}^{4}} + e^{2\nu_{0}}\, d\Omega_{k} ~;
\label{generalback}
\ee
the dilaton $\phi =\phi_{0}$ is a constant. Let us now focus on the 5d model,
postponing a brief analysis of the 6d one to the end of this section.

\subsection{QCD in the conformal window}

In the 5d model ``dual to QCD in the conformal window'', $k=0$ and the
solution (in $\alpha'=1$ units) has the same radius, dilaton and five
form field strength as the zero temperature one \cite{paris}
\bea
R^{2} &=& \frac{200}{50 + 7\rho^2 - \rho \sqrt{200 + 49 \rho^2}} ~,
\label{5drad} \\ [1ex]
e^{\phi_{0}} &=& \frac{\sqrt{200 + 49 \rho^2} - 7\rho}{10 Q_{c}} ~,
\label{5ddil} \\ [2ex]
F_{(5)} &=& Q_c\, \vol(AdS) ~,
\eea
where
\be
\rho\equiv \frac{Q_f}{Q_c}\sim \frac{N_f}{N_c} ~.
\label{defrho}
\ee
$Q_c$ and  $Q_f$ are constants proportional to $N_c$ and $N_f$ respectively.
\footnote{If we use the same normalizations as in 10d, we have $Q_f = 2 k_5
T_4 N_f = N_f/(2\pi)^2$, and $Q_c = 2 k_5 T_3 N_c = N_c/(2\pi)$. Even without
the form of $T_3, T_4$ but with only their ratio, one can get $Q_f/Q_c =
N_f/(2 \pi N_c)$.}

Note that both the $AdS$ radius (\ref{5drad}) and the dilaton (\ref{5ddil})
-- the latter being related to the gauge theory coupling by $g_{QCD}^2
\approx e^{\phi_0}$ -- depend on the flavor/color ratio $\rho$. Their
dependence is the opposite: while $R$ increases with $\rho$, the coupling
decreases. The latter is consistent with the known fact that the zero
temperature theory should be weakly coupled in the upper part of the
conformal window, that is when $\rho$ is the largest. Note that the
behavior of the coupling is as expected in the Veneziano limit: it is
given by ${\cal F}(\rho)/N_c$ for some function ${\cal F}$ whose
behavior is precisely ${\cal F}(\rho) \sim 1/\rho$ for large $\rho$
\cite{veneziano}.

The black hole temperature and entropy density read \footnote{Here and
in the following we use, for the d-dimensional Newton constant, the
expression $16\pi G_{d} = (2\pi)^{(d-3)} (\alpha')^{(d-2)/2} g_s^{2}$.}
\be
T = \frac{u_{H}}{\pi R^{2}} ~, \qquad\qquad s = \frac{A_{3}}{4G_{(5)}}
= \frac{\pi^{2} R^{3} T^{3}}{e^{2\phi_{0}}} ~,
\label{s5d}
\ee
where $A_{3}$ is the horizon area. 
The free energy can be obtained by suitably renormalizing the Euclidean
action \footnote{We thank J.M. Pons and P. Talavera for many discussions
on this procedure and for common work on a related subject.}
\be
I = \frac{1}{16\pi G_{(5)}} \int d^5x \sqrt{g}\left[ e^{-2\phi} \left(
R + 4 (\partial_\mu\phi)^2 + 5 \right) - \frac{1}{5!} F_{(5)}^2 -
2 Q_f e^{-\phi} \right] ~,
\label{action}
\ee
where $5=10-d$ is the constant needed to cancel the central charge. Since
the dilaton is constant, the DBI piece (the last one in (\ref{action}))
plays the role of an additional cosmological constant term and the
calculation is then practically the same as in ten dimensions
\cite{witten}. Performing it, the free energy (density) reads
\be
F = T I = - \frac{\pi^2 R^3 T^4}{4 e^{2\phi_0}} ~. 
\ee  
The energy density can be derived directly as $\epsilon = \partial_\beta
I$ ($\beta \equiv 1/T$) or from the ADM definition of the energy
\be
E = \frac{1}{8\pi G_{(5)}} \int d^3x \sqrt{|g^{00}|}(K_T - K_0) ~,
\ee
where $K_T, K_0$ are respectively the extrinsic curvatures of the 4d
constant time black hole and zero temperature backgrounds. The result
is the same:
\be
\epsilon = \frac{3 \pi^2 R^3 T^4}{4 e^{2\phi_0}} ~,
\ee
so that the thermodynamic relation \footnote{And the relation $s = -
\partial_T F$.} $F = \epsilon - T s$ is verified. The heat capacity
(density) $c_V$ and the speed of sound $v_s$ are (see, for example,
\cite{Myers})
\be
c_V = \partial_T \epsilon = \frac{3 \pi^2 R^3 T^3}{e^{2\phi_0}} ~,
\qquad\qquad v_s^2 = \frac{s}{c_V} = \frac13 ~,
\ee
the latter result being consistent with the zero temperature theory
being conformal.

The holographic evaluation of the ratio between the shear viscosity
$\eta$ and the entropy density in the plasma dual to this black hole 
gives the same universal result 
\be
\frac{\eta}{s} = \frac{1}{4\pi} ~,
\label{etas}
\ee
as for generic gauge theories equipped with gravity duals. The reason is
that this black hole solution and the effective action from which it is
deduced, have the general symmetries advocated in \cite{kovtuns} for this
result to be valid. Formula (\ref{etas}) is somehow a trivial result in
this setting, since we are limiting ourselves to the two derivative action,
ignoring $\alpha'$ corrections that are most likely going to modify the
ratio. If the latter were not the case, (\ref{etas}) would be a proof of
the fact that the ubiquitous matter in the adjoint representation of critical
string duals does not give any contribution to $\eta/s$. Notice, in this
respect, that the fact that matter in the fundamental representation does
not change this result has already been argued in the ten dimensional model
in \cite{cnp}, that supposedly accounts for the full backreaction of the
flavor branes in a SQCD--like theory. The result (\ref{etas}) is another
example of this fact. From the expressions of the entropy density given
above one can thus immediately deduce the shear viscosity
\be
\eta = \frac{\pi R^{3} T^{3}}{4 e^{2\phi_{0}}} ~.
\ee

All the thermodynamic quantities $F, \epsilon, c_V, \eta$ have the same
qualitative behavior as functions of $\rho$ as the entropy density. The
latter is a monotonically increasing function of $Q_c^2$ (for fixed
$\rho$) and $\rho$. Its asymptotics in $\rho$ are \footnote{Of course,
these expressions must be taken with a grain of salt: since the
background is expected to be corrected by order one terms, the numerical
coefficients are not trustworthy. Moreover, since the dual field theory
should be QCD in the conformal window for definite, finite values of
$\rho$, the limit $\rho \rightarrow 0$ ($\rho\rightarrow \infty$) is
meaningful only as an indication of the behavior for small (large)
$\rho$. The strict $\rho= 0$ case is the finite temperature version
of the unflavored Polyakov solution \cite{wall}, that could be 
dual to a YM theory without flavors.}
\be
s \sim 4 \pi^{2} Q_c^2 T^3 ~\Bigg\{ \begin{array}{ll} 1 + \sqrt{2} \rho
+ \dots \qquad & {\rm for}\ \rho\rightarrow 0 ~, \\ [2ex]
\frac{343}{250} \sqrt{\frac{7}{5}}\, \left( \rho^2 + {\cal O}(\rho^0)
\right) \qquad & {\rm for}\ \rho\rightarrow \infty ~. \end{array}
\label{sasymp5d}
\ee
The scalings with the number of colors $Q_c$ and the temperature $T$ are
the same as in ${\cal N}=4$, as expected. But, most importantly, this
solution exhibits a nice dependence on the number of fundamental flavors
$Q_f$. As one increases $Q_f$, $s$ increases: the flavors contribute
positively to the entropy density. At small $\rho$, (\ref{sasymp5d}),
the correction to the pure glue scaling $s\sim Q_c^2$ is the expected
one, namely $Q_c Q_f$. Moreover, noting that the gauge theory 't Hooft
coupling goes as $\lambda\equiv e^{\phi_{0}} Q_c \sim \sqrt{2} - 7\rho/10
+ \dots$ for $\rho\rightarrow 0$, it is tempting to write the first $\rho$
correction to the pure glue result as $\lambda\, Q_c Q_f$. This is the same
enhancement effect of the flavor contribution found in the probe setting
of ${\cal N}=4$ SYM theory in \cite{Myers}. On the other hand, the behavior
for large $\rho$, (\ref{sasymp5d}), is found to be $s\sim Q_f^2$ and the
first correction is already of order $Q_c^2$.

\subsection{Drag force and jet quenching}

Now we turn to the discussion of the energy loss of partons in this plasma.
As it is well known, the phenomenon of jet quenching observed at RHIC demands
for a very efficient mechanism of energy loss. Recently, two main ways of
accounting for this phenomenon have appeared in the literature. In \cite{liu},
the main physical process for the energy loss relevant at RHIC, that is gluon
radiation, is deduced in the perturbative QCD regime. All the effects of the
strongly coupled plasma are encoded in just one quantity, the jet quenching
parameter $\hat q$, that accounts for the non-perturbative repeated soft
interaction of the probe and the emitted gluons with the plasma. So, the
problem is reduced to the calculation of $\hat q$, which is postulated to
be given by the coefficient of the $L^{2}$ term  of a partially light-like
Wilson loop of dimensions $L^{-}\gg L$. The calculation of this Wilson loop
is then a straightforward exercise in string theory. This description should
be the right one for partons with large transverse momentum.

In a second approach initiated in \cite{herzog1,gubser}, the energy loss
process is treated completely at strong coupling. The parton of mass $m$,
moving with velocity $v$, is viewed as the extremum of a macroscopic string
attached to a probe flavor brane. In order to keep the velocity fixed, one
has to apply a drag force $f$, giving the parton some energy and momentum,
that are ultimately lost in the plasma during the motion. This transfer of
energy and momentum is seen as a flow of energy and momentum in the string,
from the probe brane to the horizon. The configuration and the momentum flow
of this string can be explicitly calculated, allowing to derive the drag
coefficient $\mu$ from the equation $f=\mu p$, where $p$ is the momentum.
Then, of course, the coefficient enters the equation $\dot p=-\mu p$, so
its inverse is the relaxation time of the parton in the plasma. Being the
entire process analyzed in string theory, and so at strong coupling, this
description should be suitable for heavy quarks with relatively small
transverse momentum.

In this section, instead of calculating the two quantities $\mu$ and
$\hat q$ again (as we will do in the next section), we just note that,
indeed, the calculations made for the critical $AdS_{5}\times S^5$ black
hole can be directly applied to the non-critical backgrounds. Thus, the
general expression of the drag force (using the relation $p/m =
v/\sqrt{1-v^2}$) and jet quenching parameter are \cite{herzog1,gubser,liu}
\bea
\frac{dp}{dt} &=& - \frac{\pi}{2} R^{2} T^{2}\, \frac{v}{\sqrt{1-v^{2}}}
~, \\ \hat q &=& \frac{\pi^{3/2} \Gamma(\frac34)}{\Gamma(\frac54)}\,
R^{2} T^{3} ~.
\eea
The novelty is that these quantities are now sensitive to the number of
colors and flavors in our plasma. The asymptotics for the relaxation time
are
\be
t \sim \frac{m}{2\pi T^{2}} ~\Bigg\{ \begin{array}{ll} 1 - \frac{\sqrt{2}}{5}
\rho + \dots \quad & {\rm for}\ \rho\rightarrow 0 ~, \\ [2ex]
\frac{5}{7} + {\cal O}\left(\frac{1}{\rho^2}\right) \quad & {\rm for}\
\rho\rightarrow \infty ~. \end{array}
\label{asymptimet5d}
\ee
Using as a representative temperature of the plasma $T = 250$ MeV,
the relaxation time of a charm quark (taking $m=1.4$ GeV) turns out to be
\be
t \,\sim\, 0.7,\ 0.6,\ 0.5 ~{\rm fm} \qquad {\rm for} \qquad \rho = 0,\
1,\ \infty ~,
\ee
while $t \,\sim\, 0.49, 0.40, 0.35$ fm for $T = 300$ MeV and, respectively,
the same values of $\rho$. Notice that the variation with $\rho$ is very
small and the values remain in the expected range of RHIC data (between
approximately $0.3$ and $8$ fm \cite{gubser}). 

The jet quenching parameter is a monotonically increasing function of
$\rho$. Thus, it does depend on the number of degrees of freedom once the
flavors are included. The asymptotics for the jet quenching parameter are
inversely proportional to those of the relaxation time,
\be
\hat q \sim \frac{4 \pi^{3/2} \Gamma(\frac34)}{\Gamma(\frac54)}\, T^{3}
~\Bigg\{ \begin{array}{ll} 1 + \frac{\sqrt{2}}{5} \rho + \dots \quad &
{\rm for}\ \rho\rightarrow 0 ~, \\ [2ex]
\frac{7}{5} + {\cal O}\left(\frac{1}{\rho^2}\right) \quad & {\rm for}\
\rho\rightarrow \infty ~. \end{array}
\label{asympqhat5d}
\ee
If this qualitative behavior remains true in the full $\alpha'$ corrected
background, this recasts the possibility of using $\hat q$ as a direct
measure of the temperature of the quark--gluon plasma, as advocated in
\cite{liu}. In fact, there is also a dependence on the ``effective number
of massless quarks'' $\rho$; this is ultimately a function of the
temperature too, albeit an unknown one. 
Putting in the numbers,
\be
\hat q=2.4,\ 2.9,\ 3.4 \quad {\rm GeV}^2/{\rm fm} \qquad {\rm for} \qquad
\rho=0,\ 1,\ \infty \qquad {\rm at}\quad T=250 \quad {\rm MeV} ~.  
\ee
In order to compare with \cite{liu}, at $T=300$ MeV one has 
\be
\hat q=4.1,\ 5.0,\ 5.8 \quad {\rm GeV}^2/{\rm fm} \qquad {\rm for} \qquad
\rho=0,\ 1,\ \infty \qquad {\rm at}\quad T=300 \quad {\rm MeV} ~,
\ee
so that it remains in the vicinity of the range of values compatible with
RHIC data (5-15 GeV$^2/$fm \cite{liu}) for every value of $\rho$, and it
is actually above 5.0 GeV$^2/$fm for $\rho > 1$ and $T = 300$ MeV.
Anyway, aside from the
actual numbers, the interesting fact is that, as already noted, the
variation of $\hat q$ is very small in the whole range of $\rho$. It
signals the fact that the flavor contribution is not changing drastically
the properties of the plasma. This is compatible with (and in a sense
gives a reason for) the evidence that the actual values of plasma properties
computed with gravity duals including only adjoint fields are very similar
to the experimental ones. Even if at present the numerical values should
not be taken very seriously as phenomenological predictions, their
robustness suggests that the string/gauge correspondence provides an
avenue that must be pursued.

An interesting comment is in order at this point. It was shown in \cite{liu3}
that, for plasmas having 10d duals of the form $AdS_{5}$--black hole times
Sasaki--Einstein five-manifolds, ratios of jet quenching parameters are
proportional to (the square root of) ratios of entropy densities. It was
then conjectured that this result might also hold for approximately conformal
systems such as the quark--gluon plasma of QCD. The authors of \cite{liu3},
however, ask themselves whether this relation is affected or not by the fact
that some of the degrees of freedom in QCD are fundamentals. Let us assume
that it is not affected. Then, it would be possible to extract the leading
correction in $N_f/N_c$ to the jet quenching parameter first computed in
\cite{liu} for ${\cal N}=4$ SYM theory, from the corresponding term in the
expansion of the entropy density obtained for the D3/D7 system \cite{mateosf}.
It gives
\be
\hat q \approx \hat q_{{\cal N}=4} \left( 1 + \frac{\lambda}{16\pi^2}
\, \frac{N_f}{N_c} \right) ~,
\ee
a result that qualitatively agrees with the leading behavior
(\ref{asympqhat5d}) of our non-critical string/gauge dual prediction. 
Nevertheless, note that the conformal non-critical result, if not drastically
modified by $\alpha'$ corrections, has a different behavior with respect to
the proposal in \cite{liu3} in that, say, the ratio of the jet quenching
parameter to the one of ${\cal N}=4$ SYM is not proportional to the square
root of the ratio of the respective entropy densities at generic $\rho$.

Let us point out that the holographic expression of other plasma
properties like the screening length and the diffusion coefficient
for heavy quarks, can be immediately taken from the critical $AdS$
cases too, as any other quantity which is not affected by the
transverse directions. Thus, in particular, from \cite{liu2,liu3},
we conclude that the Debye screening length in our non critical
setup will be only dependent on the temperature and the wind
velocity, as in the ${\cal N}=4$ plasma case. 
Since at weak coupling the Debye screening
length is known to depend both on the gauge coupling and on the
number of colors and flavors, we are led to conclude, as in
the ${\cal N}=4$ plasma case, that this could well be a genuine
strong coupling effect. As for the above mentioned diffusion
coefficient, it is related to the friction coefficient $\mu$ by
\cite{herzog1} $D=T/\mu m$; it is thus given by $D=2/(\pi T R^2)$,
and, then, it is a decreasing function of $\rho$ in the present 5d
setup.

Let us conclude this part by warning the reader about the reliability
of this non-critical string model. Aside from $\alpha'$ corrections, three
other features have not been established so far \cite{paris}. The first
one is a complete discussion of the stability: there are no explicitly
tachyonic modes, apart from a possible closed string tachyon, whose
behavior was not discussed. The second one is the exact spectrum of
the D3--D4 system on a linear dilaton background in five dimensions
(whose backreaction is supposed to give the $AdS_5$ solution we have
been discussing), that is not {\it a priori} guaranteed to contain
(only) quark modes. Finally, if we take for granted the ten dimensional
expressions for the mass of the open string tachyon ($m^2 = -1/2$) and
the mass/conformal dimension relation ($\Delta = 2 \pm \sqrt{4+m^2 R^2}$),
the increasing of $R^2$ with $\rho$ makes the dimension of the quark
bilinear decrease in climbing the conformal window, contrary to the {\it
a priori} expected behavior. \footnote{We are assuming here that the open
string tachyon is the gravity field dual to the quark bilinear, so that
approaching the weakly coupled, higher edge of the conformal window
its dimension should increase and tend to the free field value,
namely three.} Of course, the assumption of the validity of the
ten dimensional formulas is far from obvious. 

Despite this lack of
control, this background seems to encode in a nice way many interesting
features in a very simple setting, so we find it worthwhile to study
them in a ``phenomenologically'' inspired way, guided by a similar
spirit to that behind the ``$AdS$/QCD'' setting \cite{katz,pomarol}. In
particular, this is the first case where the explicit (non-quenched)
dependence on the number of flavors is exhibited for plasma observables
in the holographic setting.  

\subsection{SQCD in Seiberg's conformal window}

Let us briefly discuss the 6d model ``dual to SQCD'' ($k=1$ in
(\ref{generalback})), whose solution reads \cite{km}
\be
R^2 = 6 ~, \qquad e^{2 \nu_{0}} = \frac{2}{3 \rho^{2}} ~, \qquad
e^{\phi_{0}} = \frac{2}{3 Q_c \rho} ~, \qquad \chi = Q_c \theta ~,
\ee
where $\chi$ is the zero-form potential and $\theta$ is the angle
parameterizing $\Omega_1$. Notice that the $AdS$ radius $R$ is now
independent on both $Q_c, Q_f$, while the dilaton is not. As in the
5d model, the coupling $e^{\phi_0}$ has the correct behavior with
$\rho$, but this time it is elementary. 

As in the 5d model, the temperature $T$ and entropy density $s$ can be
computed,
\be
T = \frac{u_{H}}{\pi R^2} ~, \qquad\qquad s = 27 \pi^2\, Q_c^2\,
\rho\, T^3 ~,
\ee
as well as the free energy $F$ and energy densities $\epsilon$,
\be
F = - \frac{27 \pi^2\, Q_c^2\, \rho\, T^4}{4} ~, \qquad\qquad \epsilon
= \frac{81 \pi^2\, Q_c^2\, \rho\, T^4}{4} ~,
\ee
and the heat capacity $c_V$, speed of sound $v_s$ and shear viscosity
$\eta$, 
\be
c_V = 81 \pi^2\, Q_c^2\, \rho\, T^3 ~, \qquad\qquad v_s^2 = \frac13 ~,
\qquad\qquad \eta = \frac{27 \pi\, Q_c^2\, \rho\, T^3}{4} ~.
\ee
In this case, the dependence of the thermodynamics quantities on
$Q_c^2$ and $\rho$ is just linear. Instead, both the drag force
coefficient and the jet quenching parameters are independent on
$Q_{c},Q_{f}$, since $R^{2}$ is a constant. The values are again
slightly different from those in the ${\cal N}=4$ SYM theory. For
a charm quark at $T=250$ MeV, the relaxation time is 
\be
t = \frac{2m}{\pi R^2 T^2}\sim 0.5~{\rm fm} ~,
\ee
that is smaller than the ${\cal N}=4$ value but still in the allowed
range that results from RHIC data. The jet quenching parameter at the
same temperature is $\hat q=3.6$ GeV$^2/$fm. In order to compare with
the values in \cite{liu}, at $T=300$ MeV one finds $\hat q=6.2$ GeV$^2/$fm,
that is, higher than the ${\cal N}=4$ one and so more comfortably within
the RHIC range.

As in the 5d model, the screening length will be independent on the
details of the plasma. The diffusion coefficient $D$ will be
independent on $N_c, N_f$.

\section{QGP and wrapped fivebranes}
\label{fivebranes}

A one parameter family of black hole solutions in the background sourced
by $N_c$ {\it color} wrapped D5--branes and $N_f$ (smeared) {\it flavor}
D5--branes was presented in \cite{cnp}. This is conjectured to be the
thermal deformation of the string/gauge dual to an ${\cal N}=1$ SQCD with
quartic superpotential at the ``conformal point'' $N_f = 2 N_c$ ($\beta=0$,
no breaking of $U(1)_R$), coupled to Kaluza--Klein adjoint matter. The
string frame metric and dilaton read (setting $g_s=1$)
\bea\label{metric}
ds^2 &=& e^{\Phi_0} z^2\, \Bigg[-{\cal F} dt^2 + d\vec x_3^2 + N_c \alpha'
\, \Bigg( \frac{4}{z^2 {\cal F}}\, dz^2 + \frac{1}{\xi}\, (d\theta^2 +
\sin^2{\theta}\, d\varphi^2) \nonumber \\ 
&& + \frac{1}{4-\xi}\, (d\tilde\theta^2 + \sin^2{\tilde\theta}\,
d\tilde\varphi^2) + \frac14\, (d\psi+\cos{\theta}\, d\varphi +
\cos\tilde\theta\, d\tilde\varphi)^2 \Bigg) \Bigg] ~, \\ 
e^{\Phi} &=& z^2 e^{\Phi_0} ~,  \qquad\qquad {\cal F} = 1 -
\frac{z_0^4}{z^4} ~,
\eea
and $\xi$ is the parameter labelling the family. The horizon is at $z_0$;
for $z_0=0$ one recovers the zero temperature solution; $t,\ x_i$ are
dimensional, while the other coordinates are dimensionless. The
temperature and entropy density of this BH are 
\be\label{cnpentropy}
T = \frac{1}{2\pi\sqrt{\alpha' N_c}} ~, \qquad\qquad  
s = \frac{A_{8}}{4 G_{(10)}} = \frac{8 e^{4\Phi_0} z_0^8
N_c^4}{\xi (4-\xi)}\, T^{3} ~,
\ee
where $A_{8}$ is the area of the horizon. The fact that the temperature
does not depend on $z_0$ seems to be a common feature of black hole
backgrounds of this kind, obtained from NS5 and D5--brane configurations.
This also implies that $T$ is independent on the energy density and,
therefore, the free energy should be zero at this classical level.
Then, the theory is in a Hagedorn phase and, indeed,
the temperature above coincides with the Hagedorn temperature of Little
String Theory (LST). This suggests that there could be thermodynamical
instabilities in this solution, in the very same way as it happens in
the standard LST case \cite{kutasov,buchel}.

The instabilities are
related to the UV LST completion of the model and thus cannot be cured
by introducing IR cutoffs. In fact, we present in appendix
\ref{spherical} a new family of black hole solutions that should
be string/gauge duals of the above plasma compactified on $S^{3}$,
and argue that it should have analogous instabilities (as well as
analogous screening properties) as its uncompactified counterpart.
In view of these possible problems, it is not clear if these solutions
can provide reliable descriptions of 4d finite temperature properties
of string/gauge duals. Anyway, it is worth studying their features
as long as they are the only known 10d black hole backgrounds where
the backreaction of the flavor branes is taken into account.

Concerning the hydrodynamical properties of the $N_f = 2 N_c$ plasma,
it was already pointed out in \cite{cnp} that the ratio bewteen the
shear viscosity and entropy density has the universal value $1/4\pi$
in the limit where the gravity approximation can be trusted. This
result depends only on the symmetries \cite{kovtuns} of the black
hole solution and, thus, is not affected by the presence of the
flavor DBI term in the gravity action (as it happens in
\cite{mateosf}), nor by the troublesome thermodynamical behavior
mentioned above. This seems to be related to the fact that the
$\eta/s$ ratio depends on the properties of the horizon, while
the LST behavior is an UV feature.

\subsection {The $q \bar{q}$ potential}
\label{pot}

The goal of this section is to extract the potential between a probe quark
and a probe antiquark in the $N_f = 2 N_c$ QGP dual to the black hole
backgrounds presented above. As in the zero temperature case studied in
\cite{cnp}, we will find that the quark and antiquark are always screened.

The Wilson loop for these two non-dynamical quarks separated by a distance
$L$ in the gauge theory coordinates should be evaluated
\cite{Kinar1998,Sonnenschein1999} by the action of a fundamental
string that is parameterized by $t = \tau,\ x_1 = \sigma,\ z = z(\sigma)$.
Using the string frame metric \eqref{metric}, we find that the length of
this string is given by
\be
L(z_*) = 4 \sqrt{N_c \alpha'} \int^\infty_{z_*} \frac{z \sqrt{z_*^4 -
z_0^4} \, dz}{\sqrt{(z^4 - z_0^4)(z^4 - z_*^4)}} = 2 \sqrt{N_c \alpha'}
\, \sqrt{1 - \frac{z_0^4}{z_*^4}}\ {\rm K} \left[ \frac{z_0^4}{z_*^4}
\right] ~, 
\ee
where $z_*$ is the minimum value of the radial coordinate reached by the
string and $z_0$ is the position of the horizon. By K$[x]$ we denote the
complete elliptic integral of the first kind. The length is a monotonic
function that goes from a minimum value of $0$, attained at $z_*=z_0$,
to the limiting value of $\pi \sqrt{N_c\, \alpha'}$, for $z_* \rightarrow
\infty$.

The energy of the string, upon subtraction of the infinite masses of the
two quarks, reads:
\bea
E(z_*) &=& \frac{2 \sqrt{N_c}}{\pi \sqrt{\alpha'}} \left[ \int_{z_*}^\infty
\frac{e^{2\Phi_0}\, z^3\, {\cal F}^{1/2}(z)}{\sqrt{e^{2 \Phi_0} (z^4\,
{\cal F}(z) - z_*^4\, {\cal F}(z_*)})}\, dz - \int_{z_0}^\infty
e^{\Phi_0}\, z \, dz  \right] \nonumber \\ [1ex]
&=& \frac{2 \sqrt{N_c}}{\pi \sqrt{\alpha'}}\, e^{\Phi_0}\, z_*^2 
\left[ \int_1^\infty \left( \frac{x \sqrt{x^4 -
\frac{z_0^4}{z_*^4}}}{\sqrt{x^4-1}} - x \right) dx - \frac{1}{2}
\left( 1 - \frac{z_0^2}{z_*^2} \right) \right] ~.
\eea
It vanishes for $z_*=z_0\, \leftrightarrow\, L = 0$ but is otherwise
positive, which means that the quark and antiquark are always screened,
as in the zero temperature case. As a matter of fact, we will argue in
the following that total screening is a common feature of all known
quark--gluon plasmas having LST as its UV completion. It is worth asking
whether this affects or not other properties of the plasma, like the
drag force on a probe quark and the jet quenching parameter. Let us
address this question in the following subsections.

\subsection{Drag force}

Let us now evaluate the energy loss and the drag force of a heavy quark
moving through the plasma, following the general procedure described
in \cite{herzog1,gubser,herzog2}. In the string/gauge picture, the quark
is a macroscopic string with an extremum attached to a probe flavor
D5--brane placed at distance $z_{m}$ from the black hole horizon. We consider
a simple string configuration representing a test quark moving in a
given spatial direction $x$: $t = \tau,\ z = \sigma,\ x = x(\sigma,\tau)$.
The static configuration $x = {\rm constant}$ corresponds to a quark at
rest in the plasma. Its mass is given by the energy of the static string
hanging down from $z_m$ to the horizon
\be
M_{rest} = \frac{1}{2\pi \alpha'} \int^{z_m}_{z_0} \sqrt{-g_{tt}g_{zz}}
\, dz = \frac{e^{\Phi_0}\,\sqrt{N_c}}{2\pi\sqrt{\alpha'}}\,
(z_m^2-z_0^2) ~.
\ee
Thus, to get a very heavy quark we must require that the probe D5--brane
is placed very far from the horizon. 

Let us now consider the string configuration corresponding to a quark
which moves at constant velocity. This is achieved by setting
$x(\sigma,\tau) = u(\sigma) +v\,\tau$. From the Nambu--Goto action on
the background (\ref{metric}), it follows that the string Lagrangian
density is proportional to
\be
{\cal L} = e^{\Phi_0}\, \sqrt{4 N_c\,\alpha'\, z^2 + z^4\, {\cal F}(z)\,
u'^2 - 4 N_c\,\alpha'\, v^2\, \frac{z^2}{{\cal F}(z)}} ~,
\ee
from which it follows that the canonical momentum density conjugate to
$u(\sigma)$ is
\be
\pi_{u} = \frac{e^{\Phi_0}\, z^4\, {\cal F}(z)\, u'}{\sqrt{4
N_c\,\alpha'\, z^2 + z^4\, {\cal F}(z)\, u'^2 - 4 N_c\,\alpha'\,
v^2\, \frac{z^2}{{\cal F}(z)}}} \equiv A ~,
\ee
and, finally,
\be
u' = \frac{\sqrt{4 N_c\,\alpha'\, A^2}}{z^4 - z_0^4}\, \sqrt{\frac{z^2\,
(z^4 - z_0^4 - v^2\, z^4)}{e^{2\Phi_0}\, (z^4 - z_0^4) - A^2}} ~.
\ee
For the above square-root to be real, at the point $z_c= z_0/(1-v^2)^{1/4}$
at which the numerator changes sign, also the denominator must change sign,
this leading to 
\be
A = \frac{e^{\Phi_0}\, z_0^4\, v^2}{\sqrt{1 - v^2}} ~.
\ee
The momentum loss of the steady moving quark is thus given by
\be
\frac{dp}{dt} = \frac{e^{\Phi_0}\, z_0^2\, v}{2\pi\,\alpha' \sqrt{1 -
v^2}} ~.
\ee
Assuming now that the following relation holds, at least for low velocity,
\be
\frac{dp}{dt} = -\mu p ~,
\ee 
and that we can trust the relativistic relation
\be
p = \frac{M_{kin} v}{\sqrt{1 - v^2}} ~,
\ee
where $M_{kin}$ is the kinetic mass of the quark, one obtains the
coefficient
\be
\mu M_{kin} = \frac{e^{\Phi_0}\,z_0^2}{2\pi\,\alpha'} = 2\pi\,
e^{\Phi_0}\, z_0^2\, N_c\, T^2 ~,
\ee
or, equivalently, the relaxation time $t=1/\mu$. Let us notice that
$e^{\Phi_0}\, z_0^2$ is the value of the dilaton at the horizon,
$e^{\Phi}_0$. In the unflavored wrapped D5--brane setup of \cite{mn},
the value of the dilaton at the minimal radial position $e^{\Phi}_0$
multiplied by $N_c$ gives the parameter $\lambda\equiv e^{\Phi}_0\, N_c$
that measures the ratio of the string tension with the KK masses.
This is the parameter that shall be fixed in the large $N_c$ limit,
and must be much larger than one in order for the supergravity
approximation to be reliable. Introducing also in our context
\footnote{Here $\lambda$ has not a clear interpretation as some cut-off
coupling.} the parameter $\lambda\equiv e^{\Phi_0}\, z_0^2\, N_c$, we
can write 
\be
\mu M_{kin} = 2\pi\, \lambda\, T^2 ~.
\label{mumkin}
\ee
This is independent on the velocity and scales like $T^2$, just as
it happens in the ${\cal N}=4$ SYM plasma \cite{herzog1,gubser}. The
only difference with the latter comes from the dependence on
$\lambda$, which is enhanced from $\lambda^{1/2}$ to $\lambda$
(this is a usual phenomenon in passing from $AdS$ to more general
backgrounds with three-form RR fluxes). 

It thus seems that the energy loss of a probe quark moving in the QGP
dual
to (\ref{metric}) is not very much affected by the presence of dynamical
massless quarks.

Let us end this section by noticing that in order to have the friction
coefficient $\mu$ disjointed from the kinematical mass $M_{kin}$, we
should analyze, as in \cite{herzog1}, the small fluctuations around
the trivial static configuration $x = {\rm const}.$, assuming then a
time dependence of the type $x \sim e^{-\mu t}$. We do not pursue this
analysis here. Instead we just notice that the limiting value of $\mu$,
{\it i.e.}, the one extrapolated in the extremal case where the probe
brane touches the horizon $z_m\rightarrow z_0$, is expected to be as
the one advocated in \cite{herzog2} for general asymptotically $AdS$
backgrounds. This can be easily deduced from the general results in
\cite{antonyan}. For a metric of the form
\be
ds^2 = - \Omega(r)^2\, \Delta(r)^2\, dt^2 + \Omega(r)^2\, \left[
g_{\mu\nu}\, dx^\mu\, dx^\nu \right] + \Omega(r)^2\, \left[ g_{rr}\,
dr^2 + g_{ij}\, dy^i\, dy^j \right] ~,
\ee
where $\mu,\nu = 1, \ldots, d;\, i,j = d+2, \ldots, D$, the limiting
value of $\mu$ is given by 
\be
\mu_* = \frac{\Delta}{2\, \sqrt{g_{rr}}}\, \frac{\partial_r \Omega}{\Omega
- \Omega_0}\Bigg|_{r = r_0} ~.
\ee
Applying this formula to the $N_f = 2 N_c$ black hole, we find
$\mu_* = 2 \pi T$ as in the asymptotically $AdS$ cases.
\footnote{Incidentally, we notice that the same result is shared by
all the other known (wrapped) fivebrane black hole solutions we will
refer to in section \ref{gen}.}

\subsection{Jet quenching parameter} 

Let us now evaluate the jet quenching parameter in the $N_f = 2 N_c$ QGP
under study, following the prescriptions given in \cite{liu}. What we need
to compute is a rectangular light-like Wilson loop with one of the sides 
having a much smaller length, $L$, than that of the light-like side, $L^-$.
The prescription is that the Wilson loop in the fundamental representation
(approximated by the renormalized Nambu--Goto action $S_{NG,r}$) is related to the 
jet quenching parameter $\hat q$, according to the relation
\be
\hat q = \frac{8\sqrt{2}}{L^- L^2}\, S_{NG,r} ~.
\label{quenching}
\ee
Let us first rewrite the relevant part of the metric in light-cone
coordinates $x^{\pm} = (t\pm x_1)/\sqrt{2}$:
\bea
ds^2 &=& e^{\Phi_0} z^2 \Bigg[ - ({\cal F}(z) + 1)\, dx^+ dx^- + \frac12
(1 - {\cal F}(z))\, ((dx^+)^2 + (dx^-)^2) \nonumber \\ [1ex]
&& + d x_2^2 + d x_3^2 + N_c\,\alpha' \Bigg( \frac{4}{z^2\, {\cal
F}(z)}\, dz^2 + \dots \Bigg) \Bigg] ~.  
\eea
The string configuration for the loop is just $x^- = \tau,\ x_2 = \sigma,\
z = z(\sigma)$, with the boundary conditions $z(\pm L/2) = \infty$. The
Nambu--Goto action for a light-like length $L^-$ reads, then,
\be
S = \frac{2 L^- e^{\Phi_0}\, z_0^2}{2 \sqrt{2}\pi\,\alpha'} \int^{L/2}_0
\sqrt{1+\frac{4\alpha' N_c\, z'^2}{z^2\, {\cal F}(z)}}\, dz ~.
\ee
The equation of motion for $x_2$ then gives
\be
z' = \sqrt{\frac{z^2\, {\cal F}(z)\, \gamma^2}{4 N_c\,\alpha'}} ~,
\ee
where $\gamma$ is an integration constant, that gives
\be
\frac{\gamma\, L}{4\sqrt{N_c\,\alpha'}} = I \equiv \int^{\infty}_{z_0}
\frac{z\, dz}{\sqrt{z^4 - z_0^4}} ~.
\ee
In order to evaluate the renormalized action, we have to subtract to this
configuration the one corresponding to two separated strings extending
from the boundary to the horizon:
\be
S_0 = \frac{2 L^-}{2\pi\,\alpha'} \int^{\infty}_{z_0} dz\,
\sqrt{g_{--} g_{zz}} = \frac{2 L^- e^{\Phi_0}\, z_0^2}{2 \sqrt{2}\pi
\,\alpha'}\, \sqrt{4 N_c\,\alpha'} \cdot I ~.
\ee
We see that at small $L$, that is, large $\gamma$, we simply obtain
$S - S_0 = 0$, so that the jet quenching parameter $\hat q$ is zero \footnote{This result was obtained in collaboration with C. N\'u\~nez.}:
\be
S_{NG,r} = S - S_0 = \frac{2 L^- e^{\Phi_0}\, z_0^2}{2 \sqrt{2}\pi
\,\alpha'} \left[ \frac{L}{2}\, \sqrt{1 + \left( \frac{2 \sqrt{4
\,\alpha'\, N_c\, z'^2}\, I}{L} \right)^2} - \sqrt{4 N_c\,\alpha'}
\cdot I\right] \rightarrow 0 ~.
\ee
As we will point out in what follows, this is a general feature in gauge
theory plasmas having Little String Theory as their UV completion. The
$\hat q = 0$ result thus shares the same universality of the screening
properties in LST plasmas. As such, these properties seem to be much
more related to the 6d non local UV theories than to the IR local ones.

\subsection{On the general behavior of ``LST plasmas''}
\label{gen}

A possible puzzling result has emerged in studying the properties of
the $N_f = 2 N_c$ plasma obtained from a system of wrapped fivebranes:
while the energy loss of a probe quark in the plasma, as described by
the drag force calculation, has been found to be non zero, the jet
quenching parameter turns out to be zero. It is known that the two
quantities are related to the energy loss of a parton in two opposite
regimes of transverse momentum (large momentum for the quenching
parameter and small momentum for the drag coefficient), but obtaining
such completely different results is an apparent puzzle. So, let us
discuss this issue a bit.

It is known that LST has a critical behavior at finite temperature:
$T = T_H$ (Hagedorn temperature) and negative specific heat characterize
the BH solution related to unwrapped NS5--branes \cite{kutasov}. The
same happens \cite{buchel} to certain black hole backgrounds studied
as finite temperature extensions of the wrapped D5--brane solution
proposed in \cite{mn} as a dual to ${\cal N}=1$ SYM theory (the MN
solution). It is not difficult to realize
that both in the flat LST case and in the wrapped setups, the drag
force is non zero, while ${\hat q}=0$. This also happens in the
thermalized version \cite{Bertoldi:2002ks} of the 3d setup studied
in \cite{schve,mna}. 

Moreover there are more general BH versions of the unflavored MN
solution, which were studied in \cite{gtv}. There, a family of finite
energy BH was found, whose temperature is above the Hagedorn temperature
of LST and that basically have the same problems. Again, zero jet
quenching and non zero drag force are expected to characterize these
solutions. Checking the occurrence of total screening is not an
immediate task, mainly due to the lack of analytic control on the
solutions in \cite{gtv}. As we show in appendix \ref{gtv}, which can
be taken as a reference for analogous calculations in general LST
cases, these solutions manifest total screening too. As it transpires
from the calculations in appendix \ref{gtv}, it is expected that the
same should happen for possible $N_f \neq 2 N_c$ plasmas in the framework
of \cite{cnp}, even though the correponding gravity backgrounds are
not currently known. A possible way to evade the above critical
behavior could have been to put the black hole on $S^3$. In this
manner, another scale is introduced by hand into the system, and it
might be expected that this allows to tune the temperature of the
system away from the Hagedorn point. As we show in appendix
\ref{spherical}, this is actually not the case and these solutions
behave in the same way as the other ``LST backgrounds''. 

These are indications that our ``jet quenching puzzle'' is not related
to the presence of dynamical quarks in the 10d backgrounds.
\footnote{There are phenomenological $AdS$/QCD models like
\cite{hep-ph/0609254} where dynamical quarks are argued to be present
and responsible for the breaking of the flux tube connecting two probe
quarks beyond some critical distance. The drag force and jet quenching
parameter for those models were found to be both different from zero
\cite{hep-ph/0608274} with the jet quenching parameter going like $T^4$.
Together with the results of Section \ref{noncrit}, this indicates
that dynamical quarks seem to create no specific problem to the
evaluation of $\hat q$; a difference of \cite{hep-ph/0609254} w.r.t.
our 10d case is that we have total screening while, in the former
paper, the screening only happens above a certain energy.}
Instead, let us observe that the jet quenching parameter is definitely
dependent on the UV behavior of the dual backgrounds, and so on the 6d
LST asymptotics, while the same dependence for the drag force is not
stringent (indeed we could consider a probe flavor brane ``not so far''
from the horizon, so to cut off the UV zone). 

This fact suggests that the result $\hat q = 0$ has to be associated
more to non-local LST modes than to their claimed local counterparts.
Analogous conclusions can be reached for the screening properties of
the wrapped fivebrane plasmas, which again appear to be universal. In fact,
the total screening result is also sensitive to the UV physics, and so
to the LST regime. In conclusion, it seems reasonable to argue that
quantities crucially sensitive to the UV behavior of the metric are
determined by LST modes. These give total screening and, as a
consequence, zero jet quenching parameter. Thus, these backgrounds
do not seem to be useful in order to study UV properties of realistic
plasmas.

On the other hand, there are properties that seem not to be affected
by almost any feature of the plasmas. For example, the $\eta/s$ ratio
has the usual form because it is determined by the universal properties
of the horizon. Analogously, the result of the drag force calculations
indicates that the form of $\mu\, M_{kin}$ is exactly the same in all
the LST plasmas and practically the same as in ${\cal N}=4$. It is
insensitive to the IR (and UV) details of the dual theory \footnote{It
is not sensitive to the fact that the UV LST completion is a non-local
theory.} and it is not substantially altered by the presence of flavors.
In this respect, it seems that the indications it gives for the energy
loss process in the dual field theory are very general and universal.
The friction that a heavy parton experiences in a strongly coupled
plasma is practically always the same, irrespectively of the features
of the dual field theory. 

\section{Concluding comments}
\label{conclusions}

In this paper we have analyzed some properties of the known backgrounds
dual to finite temperature 4d gauge theories with a large number of
flavors $N_f$ and colors $N_c$, in the limit where their ratio $\rho
= N_f/N_c$ is kept constant and different from zero. Our aim was to
obtain information on the dependence of plasma features on the number
of flavors, in order to better understand the connection of the string
theory calculations with RHIC physics.

In particular, we have studied plasma features of non-critical gravity
backgrounds conjectured to be dual to finite temperature QCD and SQCD
in the conformal window \cite{km,paris,cps}. We have analyzed the
dependence on $\rho$ of thermodynamic quantities such as the entropy,
energy and free energy density and of the jet quenching parameter and
drag coefficient for a probe parton moving in the dual plasma.
While in the ``SQCD dual'' the last two features are independent on
$\rho$, in the ``QCD dual'' all the above quantities have been shown
to be monotonically increasing functions of $\rho$. For the case of
the jet quenching parameter, this seems to refine the possibility of
using it as a direct measure of the temperature of the RHIC plasma,
since it depends on the effective degrees of freedom (that, in turn,
are also a function of $T$, albeit an unknown one). On the other hand,
as a general feature, we have found that the dependence of all the
observables on $\rho$ is quite mild, so that the numerical results
for the jet quenching parameter and the relaxation time are always
very similar to the ${\cal N}=4$ ones. This provides a first
explanation of why the latter are so similar to the experimental
results at RHIC. As far as we know, our results are the first
example of unquenched effects of flavors in finite temperature
holographic duals of gauge theories. 

As a second class of backgrounds, we have studied the 10d solution
argued to be dual to a SQCD-like theory with $N_f=2N_c$ flavors found
in \cite{cnp}. Its main features are that it exhibits total screening,
the jet quenching parameter is zero and the drag coefficient is non
vanishing. We have argued that these properties are common to all the
finite temperature theories having LST as UV completion. This fact
seems to force the interpretation that all these solutions in the UV
give a description of the LST physics and so should not be reliable
as realistic plasma duals in this regime. On the other hand, the
expression of the drag coefficient turns out to be very similar to
the ${\cal N}=4$ one, underlying again the fact that dynamical quarks
seem not to change drastically the physics of the plasmas.

Both settings we have analyzed have some important problems. The
non-critical one is not well under control, while the 10d one is
probably thermodynamically unstable. Nevertheless, the results they
provide are quite interesting and it should be worth investigating
other physical properties of finite $\rho$ physics using these solutions.
Of course, finding a finite temperature background which includes the
backreaction of the flavors and does not share the problems of the
above models would be the best starting point to study such properties.

\vskip 15pt
\centerline{\bf Acknowledgments}
\vskip 10pt
\noindent
It is a pleasure to thank R. Argurio, N. Armesto, R. Casero, C. Hoyos, L. Martucci,
D. Mateos, C. N\'u\~nez, A. Paredes, J.M. Pons and P. Talavera for useful discussions and comments on the manuscript.
The work of F. B. is supported by the Belgian Fonds de la Recherche
Fondamentale Collective (grant 2.4655.07), and also by the Belgian Institut
Interuniversitaire des Sciences Nucl\'eaires (grant 4.4505.86),
the Interuniversity Attraction Poles Programme (Belgian Science
Policy) and by the European Commission FP6 programme
MRTN-CT-2004 v-005104 (in association with V.\ U.\ Brussels).
The work of J. D. E. is supported in part by MCyT and FEDER (grant FPA2005-00188), and Xunta de Galicia (Conseller\'\i a de Educaci\'on and
grant PGIDIT06PXIB206185PR). He is a {\it Ram\'on y Cajal}
Research Fellow.
This work is partially  supported by the European
Commission contracts MRTN-CT-2004-005104, MRTN-CT-2004-503369,
CYT FPA 2004-04582-C02-01, CIRIT GC 2001SGR-00065, MEIF-CT-2006-024173.
Institutional support to the Centro de Estudios Cient\'\i ficos (CECS) from
Empresas CMPC is gratefully acknowledged. CECS is funded in part by grants
from Millennium Science Initiative, Fundaci\'on Andes and the
Tinker Foundation.

\appendix

\setcounter{equation}{0}
\renewcommand{\theequation}{\Alph{section}.\arabic{equation}}

\section {The black hole solution with spherical horizon}
\label{spherical}

In this section, we will introduce certain black hole metrics which
generalize the solutions found in \cite{cnp} for $N_f =2N_c$,
Eq. \eqref{metric}. These black holes are characterized by an $S^3$
horizon and their temperature is always greater than the Hagedorn
temperature of LST. 

The string frame metric of the compactified black hole solutions is 
\bear
ds^2&=&e^{\Phi} \Big[-{\cal F}dt^2 + R^2 d \Omega_3^2 + 
\frac{R^2 N_c \alpha'}{R^2 + N_c\alpha'}  
{\cal F}^{-1} dr^2 + N_c \alpha' \Big( \frac{1}{\xi} ( d\th^2+\sin^2\th d\varphi^2)+\rc
&&+\frac{1}{4-\xi}(d\tilde\th^2+\sin^2\tilde\th d\tilde\varphi^2)
+\frac14 (d\psi + \cos \theta d\varphi + \cos \tilde \th d\tilde\varphi)^2
\Big)\Big] ~,
\label{simplebh}
\eear
with
\be
{\cal F}= 1- e^{2r_0 - 2r} ~,
\ee
so the horizon is placed at $r=r_0$. There is also a corresponding extremal solution 
which is recovered by setting ${\cal F}=1$. 
The $S_3$ metric is written as
$$
d \Omega_3^2 = dx_1^2 + \sin^2 x_1 dx_2^2 + 
\sin^2 x_1 \sin^2 x_2 \,dx_3^2 \ .
$$
The dilaton is linear
\be
\Phi = \Phi_0 + r
\ee
and $F_{(3)}$ is given by
\bea
F_{(3)} &=& N_h \sin^2 x_1 \sin x_2 \,dx_1 \wedge dx_2 \wedge dx_3
\nonumber \\ [1ex]
&-& \frac{N_c}{4}\left(\sin\tilde\theta\,d\tilde\th \wedge d\tilde\varphi
+ \sin\theta\, d\th \wedge d\varphi \right) \wedge (d\psi 
+ \cos\theta\, d\varphi + \cos\tilde\th\, d\tilde\varphi) ~.
\eea
The radius $R$ of the $S_3$ horizon is determined by the 
flux through it 
\be
R^2 = \frac{N_h \alpha'}{2} \ .
\ee
Since the flux $N_h$ is quantized, the radius of the horizon 
takes only discrete values. By taking the formal limit $R \to \infty$ of the above metric
we find the solution of \cite{cnp}. 

The temperature and entropy density of the above black holes are
\be
T(R) = \frac{1}{2 \pi \sqrt{N_c \alpha'}} \, \sqrt{1 + \frac{N_c
\alpha'}{R^2} } = T_0 \ \sqrt{1 + \frac{N_c \alpha'}{R^2} } \ > \ T_0 ~,
\label{TR}
\ee
\be
s = \frac{A_8}{4 G_{(10)}} = \frac{8 e^{4 \Phi_0} N_c^4}{ \xi(4-\xi)}
R^3 \left( \frac{R^2}{R^2 + N_c \alpha'} \right)^{3/2} T^3(R) ~, 
\ee
where $T_0$ denotes the temperature of the black hole in \cite{cnp}.
Note that the temperature increases as the radius $R$ of the horizon
decreases.
There is a hint that the above black holes may be thermodynamically unstable.
Let us consider the above family of solutions in the limit $R \to \infty$
and assume that they are a good approximation to black holes with a flat
${\bf R}^3$ horizon like the solution in \cite{cnp}. Then, as the black hole temperature
increases as $R$ decreases, \eqref{TR}, we can see that the entropy $s \sim R^3$ 
is a decreasing function of the temperature, which is a sign of thermodynamical
instability.  
 
\subsection {The $q \bar{q}$ potential}

In this section we calculate the potential between a
probe quark and a probe antiquark in the above black hole backgrounds, as we have done in Section \ref{pot}. 
The Wilson loop for two non-dynamical quarks separated by a distance $L$
should be evaluated by the action of a fundamental
string that is parameterized by $t = \tau, x_1 = \sigma, r = r(\sigma)$.
Using the string frame metric 
\bear
ds^2&=&e^{\Phi} \Big[-{\cal F}dt^2 + R^2 d \Omega_3^2 + 
\frac{R^2 N_c\alpha'}{R^2 + N_c\alpha'}  
{\cal F}^{-1} dr^2 + N_c\alpha' \Big( \frac{1}{\xi} ( d\th^2+\sin^2\th d\varphi^2)+\rc
&&
+\frac{1}{4-\xi}(d\tilde\th^2+\sin^2\tilde\th d\tilde\varphi^2)
+\frac14 (d\psi + \cos \theta d\varphi + \cos \tilde \th d\tilde\varphi)^2
\Big)\Big],
\label{simplebhstring}
\eear
we find that the length $L$ and the energy of this string upon subtraction
of the infinite masses of the two quarks are given by
\begin{equation}
L(r_*) = 2 \frac{\sqrt{\gamma}}{R} \int^\infty_{r_*}
\frac{dr}{\sqrt{{\cal F}(r)}} \frac{e^{\Phi_*} {\cal F}_*^{1/2} }{\sqrt{e^{2 \Phi} {\cal F}(r) 
- e^{2 \Phi_*} {\cal F}_*}}
= 2 \frac{\sqrt{\gamma}}{R} \sqrt{ 1 - e^{2 r_0 - 2 r_*} } 
\ {\rm K}[ e^{2 r_0 - 2 r_*} ] ,
\end{equation}
\bea
E(r_*) &=& \frac{\sqrt{\gamma}}{\pi \alpha'}\, \left[ \int_{r_*}^\infty
\frac{e^{2\Phi(r)}\, {\cal F}^{1/2}(r)}{\sqrt{e^{2 \Phi(r)}\, {\cal F}(r)
- e^{2 \Phi_*}\, {\cal F}_*}}\, dr - \int_{r_0}^\infty e^{\Phi(r)}\, dr
\right] \\ [2ex]
&=& \frac{\sqrt{\gamma}}{\pi \alpha'}\, e^{\Phi_*} \left[ \int_0^1
\frac{\sqrt{1 - e^{2r_0-2r_*}\, z^2} - \sqrt{1 - z^2}}{z^2\,
\sqrt{1-z^2}}\, dz - ( 1 - e^{r_0-r_*} ) \right] ~, \nonumber 
\eea
where $r_*$ is the minimum value of the radial coordinate reached by the string and $\gamma = \frac{R^2 N_c\alpha'}{R^2 + N_c\alpha'}$.

The length $L(r_*)$ is a monotonic function that goes from a minimum value of $0$
attained at $r_*=r_0$ to the limit value of $\frac{\pi \sqrt{\gamma}}{R}$ for 
$r_* \rightarrow \infty$. \footnote{For these black hole solutions with spherical horizon, 
$x_1$ is an equatorial angle and $L$ is then the angular separation between
quark and antiquark along the equator. 
Note also that the limiting value of $L$ is also less than $\pi$.}
The energy $E(r_*)$ vanishes for $r_*=r_0 \leftrightarrow L = 0$ but is 
otherwise positive, which means that the quark and antiquark are always screened,
as for the black holes \eqref{metric}.

\section{Total screening in GTV black holes}
\label{gtv}

In this appendix, we will study the quark-antiquark potential
in the black hole backgrounds found in \cite{gtv}.
We will find that, as in the case of the $N_f = 2N_c$ 
black hole, quark and antiquarks are always totally screened.
Furthermore, this seems to be a generic feature of black hole backgrounds associated to systems
of wrapped fivebranes. Indeed, one can repeat the analysis for the finite temperature
solutions found in \cite{Bertoldi:2002ks} and find the same result.
As it transpires from the analysis below, this is related to the asymptotics of the solution
both close to the horizon and far away from it.

In the notation where the string frame metric reads
\be
ds^2_{string} = - G_{00}(s) dt^2 + G_{x_{||}x_{||}} dx_{||}^2 + G_{ss}(s) ds^2 +
G_{x_T x_T} (s) dx_{T}^2,
\ee
define
\bea
f^2(s(x)) & =  & G _{00}(s(x)) G_{x_{||}x_{||}}(s(x)),  \nonumber \\
g^2(s(x)) & = & G_{00}(s(x)) G_{ss}(s(x)) .
\eea
The distance between quark and antiquark is given by
the general formula \cite{Kinar1998,Sonnenschein1999}
\be
L(s_0) = 2 \int_{s_0}^{s_1} \frac{g(s) f(s_0) \, ds}{f(s) \sqrt{f^2(s) -f^2(s_0)}}.
\label{LLL}\ee
In the case we are interested in, the string frame metric reads \cite{gtv} 
\be
ds^2 = e^{\Phi(r)} \Big[- \nu(r) dt^2 + d \vec{x}^2 + 
\nu(r)^{-1} dr^2 + R^2(r)( d\th^2+\sin^2\th d\varphi^2 )+\rc
\tilde{\epsilon}^2_1 + \tilde{\epsilon}^2_2 + 
\tilde{\epsilon}^2_3 \, \Big] ,
\label{simplebhstringgtv} 
\ee
so that
\bea
f^2(s(x)) & =  & G _{00}(s(x)) G_{x_{||}x_{||}}(s(x))  = \nu(s(x)) e^{2 \Phi(s(x))} ,\nonumber \\
g^2(s(x)) & = & G_{00}(s(x)) G_{ss}(s(x))  =  e^{2 \Phi(s(x))}  .
\eea
Close to the horizon, which we can set at $r=0$ \cite{gtv}
\be
\nu = \nu'_h \, r + {\cal O}(r^2), \qquad 
\Phi = \Phi_h + {\cal O}(r), \qquad
R = R_h + {\cal O}(r),  
\label{asymphorizon}\ee
whereas for $r \to \infty$ 
\be
\nu = 1 + {\cal O}(e^{-r}), \qquad 
\Phi = \Phi_\infty + \frac{r}{2} -\frac{1}{4} \log r + {\cal O}(r^{-2}), \qquad
R \sim \sqrt{2r}.
\label{asympinfinity}\ee
Due to the asymptotic behavior of the above functions, the integral 
\eqref{LLL} is convergent in the limit $s_1 \to \infty$.

We would like to show that $L(s_0)$ vanishes in the limit $s_0 \to 0$. To
this end, let us introduce the function
\be
\Upsilon(s) \equiv \frac{1}{\sqrt{\nu(s)\, \left( \nu(s)\, e^{2 \Phi(s)}
- \nu(s_0)\, e^{2\Phi(s_0)} \right)}} ~,
\ee
that is useful to write $L(s_0)$ in a compact form,
\be
L(s_0) = 2 \int_{s_0}^{\infty} \sqrt{\nu(s_0)}\, e^{\Phi(s_0)}\,
\Upsilon(s)\, ds ~.
\label{lszero}
\ee
The function $\Upsilon(s)$ is analytic in $[1,\infty[$ and its integral
is convergent. Therefore, if we split the integral in (\ref{lszero}) into
two integrals from $s_0$ to $1$, and from $1$ to $\infty$, the latter simply
vanishes in the limit $s_0 \to 0$, as long as $\sqrt{\nu(s_0)} \sim
\sqrt{s_0}$. Then,
\bea
L(s_0) &=&  2 \int_{s_0}^{1} \sqrt{\nu(s_0)}\, e^{\Phi(s_0)}\,
\Upsilon(s)\, ds \nonumber \\
&\sim& \frac{2}{\sqrt{\nu'_h}}\, \sqrt{s_0}\,
\int_{s_0}^1 \frac{ds}{\sqrt{s (s - s_0)}} \sim - \sqrt{s_0}\,
\log s_0 ~.
\eea
In summary, in the limit $s_0 \rightarrow 0$,
\be
L(s_0) \sim - \sqrt{s_0} \, \log s_0 \rightarrow 0 ~.
\ee 
This behavior of the length functional is the same for all black holes
with LST asymptotics \cite{gtv,Bertoldi:2002ks} and differs from other
more familiar cases where $L(s_0)$ actually diverges in the same limit.
It is also drastically different from the behavior in the zero temperature
solutions \cite{mn,mna} where $L(s_0)$ actually diverges and there is
linear confinement.

We will now study the energy of the Wilson loop in the limit $s_0 \to 0$ 
and show that it is positive in a neighborhood of zero. 
The energy of the regularized Wilson loop, which means that we are subtracting the masses
of the quark and anti-quark, is given by the following formula
\be
E(s_0) = 2 \int_{s_0}^{\infty} 
\left( \frac{f(s)}{\sqrt{f(s)^2 -f(s_0)^2}} - 1 \right) g(s) \, ds - 2 \int_0^{s_0} g(s) \,ds.
\label{EEE}
\ee
Given the asymptotics at infinity of the functions involved \eqref{asympinfinity}, 
the above integral is actually finite.
Since the first integrand is positive 
\be
E(s_0) > \tilde E(s_0) \equiv 2 \int_{s_0}^{2 s_0} \left(
\frac{f(s)}{\sqrt{f(s)^2 - f(s_0)^2}} - 1 \right) g(s) \, ds - 2
\int_0^{s_0} g(s) \,ds ~.
\ee
Thus, in the limit $s_0 \to 0$,
\bea
\tilde E(s_0) &=& 2 \int_{s_0}^{2 s_0} \left( \frac{f(s)}{\sqrt{f(s)^2
- f(s_0)^2}} - 1 \right) g(s)\, ds - 2 \int_0^{s_0} g(s)\, ds \nonumber
\\ [1ex]
&=& 2 \int_{s_0}^{2 s_0} \left( \nu(s)\, \Upsilon(s)\, e^{\Phi(s)} - 1
\right)\, e^{\Phi(s)}\, ds - 2 \int_0^{s_0} e^{\Phi(s)} \,ds \nonumber
\\ [1ex]
&=& 2 s_0\, e^{\Phi(s_0)} \int_{1}^{2} \left( \frac{\sqrt{t}}{\sqrt{t - 1}}
- 1 \right)\, dt\ - \ 2 s_0\, e^{\Phi(s_0)}\, + {\cal O}(s_0^2) \nonumber
\\ [1ex]
&=& \left( 2 \sqrt{2} + 2 \log ( 1 + \sqrt{2} ) - 4 \right)\, s_0\,
e^{\Phi(s_0)} + {\cal O}(s_0^{2}) \nonumber \\ [1ex]
&\approx& 0.591174\, s_0\, e^{\Phi(s_0)}\, > 0 ~.
\eea

The above is not enough to show that for this class of black holes quark
anti-quark pairs are totally screened. In fact, it could be possible that
$L(s_0)$ vanishes again and that around this value of $s_0$ the energy
is actually negative. However, numerical analysis shows that the length
$L$ is approximately a monotonic increasing function and does not vanish
again. Furthermore, one can show that in the limit $s_0 \to \infty$ 
$L(s_0)$ goes to a constant value determined by the slope of the dilaton
\be
L(s_0) \rightarrow \frac{\pi}{Z} \quad \leftrightarrow \quad \Phi(s)
\sim Z\, s \, \qquad s \to \infty ~,
\ee
as we found in the $N_f = 2N_c$ black hole. Numerical analysis also
confirms that the energy of the Wilson loops is always positive for
$s_0 > 0$.

It would be interesting to check explicitly whether black hole solutions
for $N_f \ne 2N_c$ exhibit total screening as well. This fact seems to
be determined by and intertwined with the asymptotics of the solution
close to the horizon \eqref{asymphorizon} and also the large $r$
asymptotics \eqref{asympinfinity}. 
The fact that the energy of the Wilson loop is positive for small enough
$L$ and that the length presumably does not vanish again are strong hints
that total screening might be a feature of general fivebrane or LST black
holes. Indeed, one can perform the same analysis for the black holes 
found in \cite{Bertoldi:2002ks} that have essentially the same asymptotic behavior as \cite{gtv} and again find total screening. 


\end{document}